\newcommand{\CLASSINPUTtoptextmargin}{19mm}%
\newcommand{\CLASSINPUTbottomtextmargin}{43mm}%
\newcommand{\CLASSINPUTinnersidemargin}{12.9mm}%
\newcommand{\CLASSINPUToutersidemargin}{12.9mm}%
\documentclass[conference,10pt,a4paper]{IEEEtran}%CP this assumes IEEEtran V1.8+
\usepackage{amsmath}% for double integral symbol in this template
\usepackage{times}% use times font for the paper instead of default LaTeX fonts
\usepackage{graphicx}% for figures
\usepackage{multirow}% to allow multiple-row elements in tabular environment
\usepackage[none]{hyphenat}% turn off hyphenation to make text extraction and indexing easier
\usepackage{float}% better control of floating figures and tables
\usepackage{subfig}% for labelling subfigures within figures
\usepackage{iftex}% for detecting whether XeLaTeX or pdfLaTeX is being used, for the purpose of choosing the right fonts below.
\usepackage[backend=bibtex,sorting=none,style=ieee]{biblatex}
\ifPDFTeX%
\usepackage{t1enc}% allows access to various special characters
\usepackage{times}% change font to Nimbus Roman (based on Times Roman)
\fi%
\ifXeTeX%
\usepackage{fontspec}% needed to change main font
\fi%
\ifLuaTeX%
\usepackage{fontspec}% needed to change main font
\fi%
\makeatletter

\def\@maketitle{\newpage
\bgroup\par\addvspace{0.5\baselineskip}\centering%
\ifCLASSOPTIONtechnote% technotes
   {\bfseries\large\@IEEEcompsoconly{\sffamily}\@title\par}\vskip 1.3em{\lineskip .5em\@IEEEcompsoconly{\sffamily}\@author
   \@IEEEspecialpapernotice\par{\@IEEEcompsoconly{\vskip 1.5em\relax
   \@IEEEtitleabstractindextextbox{\@IEEEtitleabstractindextext}\par
   \hfill\@IEEEcompsocdiamondline\hfill\hbox{}\par}}}\relax
\else% not a technote
   \vskip0.2em{\APMCtitlesize\ifCLASSOPTIONtransmag\bfseries\LARGE\fi\@IEEEcompsoconly{\sffamily}\@IEEEcompsocconfonly{\normalfont\normalsize\vskip 2\@IEEEnormalsizeunitybaselineskip
   \bfseries\Large}\@title\par}\vskip1.0em\par% CAUSAL PRODUCTIONS change on this line
   \ifCLASSOPTIONconference%
      {\@IEEEspecialpapernotice\mbox{}\vskip\@IEEEauthorblockconfadjspace%
       \mbox{}\hfill\begin{@IEEEauthorhalign}\@author\end{@IEEEauthorhalign}\hfill\mbox{}\par}\relax
   \else% peerreviewca, peerreview or journal
      \ifCLASSOPTIONpeerreviewca
         {\@IEEEcompsoconly{\sffamily}\@IEEEspecialpapernotice\mbox{}\vskip\@IEEEauthorblockconfadjspace%
          \mbox{}\hfill\begin{@IEEEauthorhalign}\@author\end{@IEEEauthorhalign}\hfill\mbox{}\par
          {\@IEEEcompsoconly{\vskip 1.5em\relax
           \@IEEEtitleabstractindextextbox{\@IEEEtitleabstractindextext}\par\hfill
           \@IEEEcompsocdiamondline\hfill\hbox{}\par}}}\relax
      \else% journal, peerreview or transmag
         \ifCLASSOPTIONtransmag
           {\@IEEEspecialpapernotice\mbox{}\vskip\@IEEEauthorblockconfadjspace%
            \mbox{}\hfill\begin{@IEEEauthorhalign}\@author\end{@IEEEauthorhalign}\hfill\mbox{}\par
           {\vspace{0.5\baselineskip}\relax\@IEEEtitleabstractindextextbox{\@IEEEtitleabstractindextext}\vspace{-1\baselineskip}\par}}\relax
         \else% journal or peerreview
           {\lineskip.5em\@IEEEcompsoconly{\sffamily}\sublargesize\@author\@IEEEspecialpapernotice\par
           {\@IEEEcompsoconly{\vskip 1.5em\relax
            \@IEEEtitleabstractindextextbox{\@IEEEtitleabstractindextext}\par\hfill
            \@IEEEcompsocdiamondline\hfill\hbox{}\par}}}\relax
         \fi
      \fi
   \fi
\fi\par\addvspace{0.0\baselineskip}\egroup}% CAUSAL PRODUCTIONS change on this line, reduce the vspace from 0.5\baselineskip to 0.0

\def\APMCtitlesize{\@setfontsize{\APMCtitlesize}{24}{24pt}}% CAUSAL PRODUCTIONS change on this line
\def\APMCauthorsize{\@setfontsize{\APMCauthorsize}{11}{11pt}}% CAUSAL PRODUCTIONS change on this line
\def\APMCaffilsize{\@setfontsize{\APMCaffilsize}{10}{10pt}}% CAUSAL PRODUCTIONS change on this line
\def\APMCcaptionsize{\@setfontsize{\APMCcaptionsize}{9}{10pt}}% CAUSAL PRODUCTIONS change on this line
\def\APMCbibsize{\@setfontsize{\APMCbibsize}{8}{10pt}}% CAUSAL PRODUCTIONS change on this line

\def\@IEEEauthorblockNstyle{\APMCauthorsize\@IEEEcompsocnotconfonly{\sffamily}\@IEEEcompsocconfonly{\large}}%CAUSAL PRODUCTIONS removed sublargesize to get correct APMCauthorsize
\def\@IEEEauthorblockAstyle{\APMCaffilsize\@IEEEcompsocnotconfonly{\sffamily}\@IEEEcompsocconfonly{\itshape}\@IEEEcompsocconfonly{\large}}%CAUSAL PRODUCTIONS removed normalsize to get correct APMCaffilsize
\def\@IEEEauthordefaulttextstyle{\APMCauthorsize\@IEEEcompsocnotconfonly{\sffamily}\sublargesize}%CP

\def\thebibliography#1{\section*{\refname}%
    \addcontentsline{toc}{section}{\refname}%
    \APMCbibsize\@IEEEcompsocconfonly{\small}\vskip 0.3\baselineskip plus 0.1\baselineskip minus 0.1\baselineskip% CAUSAL PRODUCTIONS change on this line
    \list{\@biblabel{\@arabic\c@enumiv}}%
    {\settowidth\labelwidth{\@biblabel{#1}}%
    \leftmargin\labelwidth
    \advance\leftmargin\labelsep\relax
    \itemsep \IEEEbibitemsep\relax
    \usecounter{enumiv}%
    \let\p@enumiv\@empty
    \renewcommand\theenumiv{\@arabic\c@enumiv}}%
    \let\@IEEElatexbibitem\bibitem%
    \def\bibitem{\@IEEEbibitemprefix\@IEEElatexbibitem}%
\def\newblock{\hskip .11em plus .33em minus .07em}%
\ifCLASSOPTIONtechnote\sloppy\clubpenalty4000\widowpenalty4000\interlinepenalty100%
\else\sloppy\clubpenalty4000\widowpenalty4000\interlinepenalty500\fi%
    \sfcode`\.=1000\relax}

\long\def\@makecaption#1#2{%
\ifx\@captype\@IEEEtablestring%
\par\@IEEEtabletopskipstrut% strut used to align table caption with facing column
\else
\@IEEEfigurecaptionsepspace
\fi
\setbox\@tempboxa\hbox{\normalfont\footnotesize {#1.}\nobreakspace\nobreakspace #2}%
\ifdim \wd\@tempboxa >\hsize%
\setbox\@tempboxa\hbox{\normalfont\footnotesize {#1.}\nobreakspace\nobreakspace}%
\parbox[t]{\hsize}{\normalfont\footnotesize\noindent\unhbox\@tempboxa#2}%
\else
\ifCLASSOPTIONconference \hbox to\hsize{\normalfont\footnotesize\hfil\box\@tempboxa\hfil}%
\else \hbox to\hsize{\normalfont\footnotesize\box\@tempboxa\hfil}%
\fi\fi
\ifx\@captype\@IEEEtablestring%
\@IEEEtablecaptionsepspace
\else
\fi}

\newlength\tablecaptiontotableskip
\newlength\figuretocaptionskip
\def\@IEEEfigurecaptionsepspace{\vskip\figuretocaptionskip\relax}%
\def\@IEEEtablecaptionsepspace{\vskip\tablecaptiontotableskip\relax}%

\def\abstract{\normalfont%
\@IEEEabskeysecsize\bfseries\textit{\abstractname}\,\bfseries\textit{---}\,%
\@IEEEgobbleleadPARNLSP}%

\def\IEEEkeywords{\normalfont%
\@IEEEabskeysecsize\bfseries\textit{\IEEEkeywordsname}\,\bfseries\textit{---}\,%
\@IEEEgobbleleadPARNLSP}%
\def\endIEEEkeywords{\relax\vspace{0.67ex}%
\par\if@twocolumn\else\endquotation\fi%
\normalsize\normalfont}%

\DeclareRobustCommand*{\APMCauthorrefmark}[1]{\raisebox{0pt}[0pt][0pt]{\textsuperscript{#1}}}%
\def\@IEEEauthorblockNtopspace{0ex}
\def\@IEEEauthorblockAtopspace{1mm}
\def\tablename{Table}% APMC: was TABLE in IEEETran, but we are now using capitalized caption text not smallcaps
\def\thetable{\arabic{table}}% APMC: Tables are numbered using arabic not roman
\def\IEEEkeywordsname{Keywords}% use Keywords instead of Index Terms
\def\subsubsection{\@startsection{subsubsection}{3}{\z@}{1.5ex plus 1.5ex minus 0.5ex}%
{0.7ex plus .5ex minus 0ex}{\normalfont\normalsize\itshape}}%
\newlength{\CPheadmatchindent}% 
\def\@seccntformat#1{\hbox to\CPheadmatchindent{\csname the#1dis\endcsname}\hskip 0.1em \relax}
\IEEEilabelindentA \parindent
\IEEEilabelindent \IEEEilabelindentA
\IEEEelabelindent \parindent
\IEEEdlabelindent \parindent
\IEEElabelindent \parindent
\makeatother

\begin{document}
%%%%%%%%%%%%%%%%%%%%%%%%%%%%%%%%%%%%%%%%%%%%%%%%%%%%%%%%%%%%%%%%%%%%%%%%%%%%%
% We use \raggedbottom to avoid latex adding vertical space around headings.
% This gives a better idea to the author about how much white space remains
% as the page limit is approached.
\raggedbottom
%
%%%%%%%%%%%%%%%%%%%%%%%%%%%%%%%%%%%%%%%%%%%%%%%%%%%%%%%%%%%%%%%%%%%%%%%%%%%%%
% PAPER TITLE AND AUTHOR BLOCK
%
% The paper title can use linebreaks \\ within to get better formatting if desired.
%
\title{
Charge-Coupled Device-Based \\Arbitrary Waveform Generation for Qubit Control
}
%
% Next we define the author names and affiliations.
%
\author{%
\IEEEauthorblockN{%
Shinichi Morisaka\APMCauthorrefmark{\#1}, 
Ryutaro Ohira\APMCauthorrefmark{\#2}
}% \IEEEauthorblockN Names
\IEEEauthorblockA{%
\APMCauthorrefmark{\#}QuEL, Inc., Japan\\
\{
\APMCauthorrefmark{1}morisaka, 
\APMCauthorrefmark{2}ohira
\}
@quel-inc.com\\
}% \IEEEauthorblockA Affils
}% \author
%
% Next we make the title/author block using the information defined above.
\maketitle
%
%%%%%%%%%%%%%%%%%%%%%%%%%%%%%%%%%%%%%%%%%%%%%%%%%%%%%%%%%%%%%%%%%%%%%%%%%%%%%
% ABSTRACT paragraph.
% in the abstract paragraph.
%
\begin{abstract}
We propose an arbitrary waveform generation architecture based on charge-coupled devices (CCDs) for qubit control. 
Waveform samples are stored as charge packets in CCD channels and sequentially clocked out as voltage samples. 
Multiple channels are operated in parallel and time-interleaved to achieve high effective sampling rates. 
The generated baseband waveforms are then upconverted to microwave frequencies to drive qubit operations. 
By representing waveform information as stored charge rather than as continuously generated or transmitted microwave signals, the proposed architecture provides a path to reducing cryogenic interconnect density and lowering heat load in scalable quantum control systems.
\end{abstract}


%%%%%%%%%%%%%%%%%%%%%%%%%%%%%%%%%%%%%%%%%%%%%%%%%%%%%%%%%%%%%%%%%%%%%%%%%%%%%

\section{
Introduction
}

Superconducting quantum processors have recently scaled to hundreds of integrated qubits~\cite{kim2023evidence, google2025quantum}. 
At this scale, delivering accurately calibrated microwave waveforms to many qubits operating in a dilution refrigerator becomes an increasingly serious engineering challenge.
In conventional architectures, these signals are generated by room-temperature electronics and routed to the qubits through coaxial lines, often with dedicated lines for individual qubits~\cite{krantz2019quantum}. 
As the number of qubits increases, each additional line consumes physical space inside the dilution refrigerator and introduces passive thermal inflow from room temperature to the cryogenic stages. 
At the same time, cryogenic components located in the cold stages contribute additional local power dissipation. 
The resulting heat load can become comparable to, or even exceed, the available cooling power, making waveform delivery itself a major bottleneck for scaling cryogenic quantum processors~\cite{krinner2019engineering}.

Several strategies have been investigated to alleviate this bottleneck. 
Frequency-division and time-division multiplexing reduce the number of physical interconnects by allowing multiple signals to share a smaller set of lines~\cite{jerger2012frequency, van2019impact, acharya2023multiplexed, ohira2024optimizing, matsuda2026selective, richter2026overhead}. 
Cryogenic control electronics, including cryogenic CMOS (cryo-CMOS), relocate part of the control functionality closer to the qubits and can reduce the need for wiring from room temperature to the cold stage~\cite{charbon2016cryo, patra2017cryo}. 
Photonic links offer another approach, transmitting information through optical fibers with low thermal conductance and converting it into electrical signals at low temperature~\cite{lecocq2021control, joshi2023scaling}.

Each of these approaches addresses an important aspect of the scaling problem, yet fundamental constraints remain. 
Multiplexed control schemes based on frequency- or time-division multiplexing are still in the early experimental stages, and their practical scalability has yet to be fully demonstrated. 
Cryo-CMOS electronics, despite considerable recent progress, still dissipate tens of milliwatts of power at the cold stage~\cite{underwood2024using}. 
Photonic links reduce passive thermal inflow by replacing coaxial cables with optical fibers, but the conversion of optical signals into microwave waveforms must still occur at cryogenic temperatures in real time. 
As a result, the heat dissipated at the cryogenic photodetector (PD) scales with the duty cycle of the control sequences: as the number of qubits and the density of gate operations increase, the PD heat load rises accordingly~\cite{lecocq2021control}. 
More broadly, in all of these architectures, the control hardware must remain active whenever waveforms are delivered, with no mechanism to separate waveform generation from heat dissipation at the cold stage.

\begin{figure}[t]
\centering
\includegraphics[width=80mm]{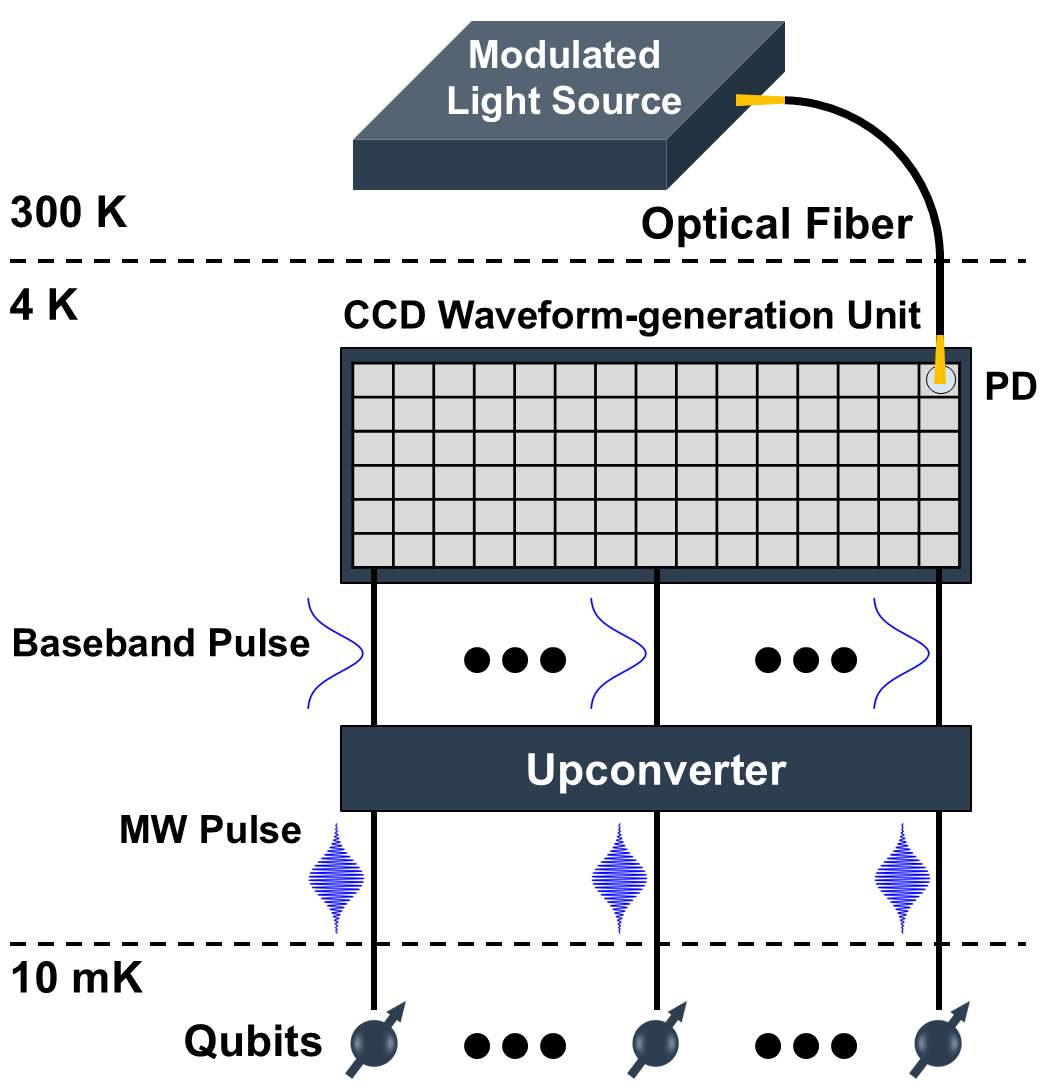}
\caption{
Proposed CCD-based architecture for qubit control.
}
\label{fig:architecture}
\vspace{-\baselineskip}
\end{figure}

The heat-dissipation problem is fundamentally linked to how waveform generation is designed: in all existing approaches, control signals are generated and delivered continuously. 
A key observation, however, is that qubit control waveforms are not inherently online: a fixed library of waveforms, including gate operations and readout pulses, is repeatedly replayed according to a sequence determined by the quantum circuit.
This enables a fundamentally different design direction: rather than continuously generating waveforms from incoming digital data, waveform information can be stored directly in analog form near the qubit and read out on demand. 

Charge-coupled devices (CCDs)~\cite{boyle1970charge, janesick1987scientific} naturally implement this store-and-readout paradigm: a CCD stores analog information as charge packets confined in potential wells, transfers these packets along a channel through sequential clock pulses, and converts them into voltage at the readout node. 
In this work, we propose an arbitrary waveform generation architecture using CCDs for qubit control, in which waveform samples are stored as charge packets in CCD channels and read out as voltage samples through sequential charge transfer and charge-to-voltage conversion. 
The generated waveform acts as a baseband signal and is upconverted using a local oscillator to produce a microwave control signal for qubit operations. 
This paper describes the system architecture, storage and readout principles, and the time-interleaving scheme of the proposed approach, which offers a distinct design direction for scalable control systems.

%%%%%%%%%%%%%%%%%%%%%%%%%%%%%%%%%%%%%%%%%%%%%%%%%%%%%%%%%%%%%%%%%%%%%%%%%%%%%

\section{
CCD-Based Waveform Generation Architecture
}

In this section, we describe the proposed CCD-based architecture for generating waveforms for qubit control.

Figure~\ref{fig:architecture} shows the overall architecture. 
The core concept is to store sampled waveform information as charge packets in CCD channels and reconstruct an analog waveform by sequentially transferring and reading out those charge packets; the CCD is used not as an image sensor but as a charge-storage and charge-transfer medium for arbitrary waveform generation.
At room temperature, a modulated light source encodes the waveform data as a pulse sequence, which is transmitted through an optical fiber to a cryogenic stage, for example, at 4~K. 
At the 4~K stage, a CCD waveform-generation unit receives the optical signal through a dedicated PD and writes the waveform data into the CCD channels via optical charge injection. 
Unlike a conventional image sensor, this application does not require a separate PD for every pixel; only a small number of PDs is needed. 
The unit then generates a baseband analog waveform, which is sent to an upconverter to produce a microwave control signal for qubit operations.

\begin{figure*}[t]
\centering
\includegraphics[width=175mm]{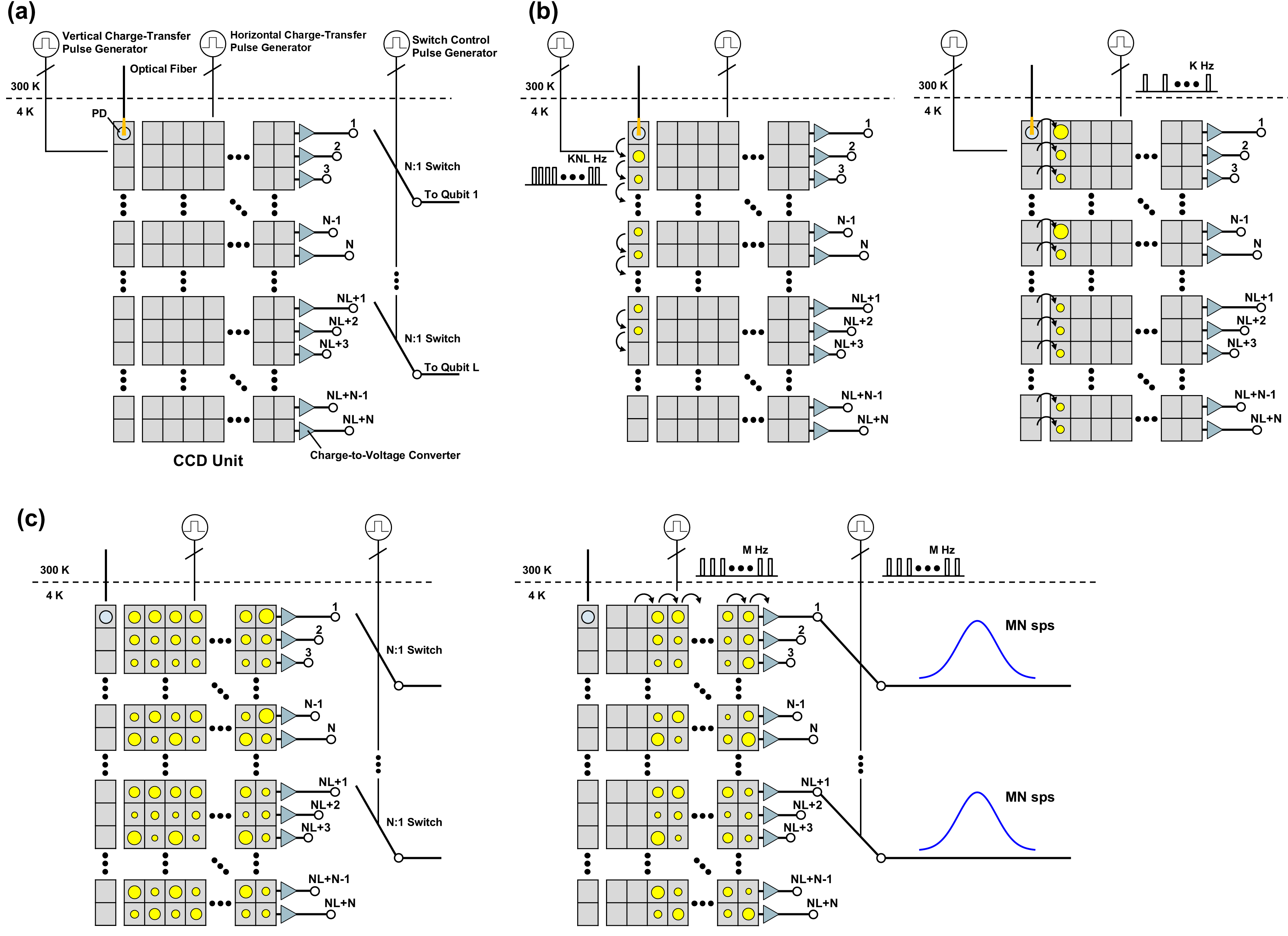}
\caption{
(a) Detailed structure of the CCD waveform-generation unit. 
(b) Waveform loading phase. 
(c) Waveform output phase.
}
\label{fig:operation}
\vspace{-\baselineskip}
\end{figure*}

Figure~\ref{fig:operation}(a) shows the detailed structure of the CCD waveform-generation unit, which consists of a CCD unit, pulse generators for vertical and horizontal charge transfer, and an $N{:}1$ switch for $L$ qubits.
The CCD unit contains multiple channels, each storing a sequence of charge packets.

Operation is divided into two phases: (1) waveform loading and (2) waveform output. 
Figure~\ref{fig:operation}(b) shows the waveform loading phase. During this phase, modulated light is directed onto the PD, injecting charge into the CCD channels in proportion to the desired waveform amplitude at each sample point. 
The size of each circle in Fig.~\ref{fig:operation}(b) represents the corresponding charge-packet amount. 
After conversion to charge at the PD, each charge packet is transferred to its designated channel and stored. 
Horizontal transfer is controlled by the horizontal transfer pulse generator at a frequency of $K$~Hz, while vertical transfer is controlled by the vertical transfer pulse generator at a frequency of $KNL$~Hz. 
Since the vertical transfer frequency is $NL$ times higher than the horizontal transfer frequency, $K$ must be set sufficiently small to keep the vertical transfer within a practical operating range for charge transfer.
As shown in the left and right panels of Fig.~\ref{fig:operation}(b), the waveform loading phase consists of repeated vertical and horizontal transfers.
First, vertical transfer is performed at $KNL$~Hz to shift all charge packets to the designated rows (left panel of Fig.~\ref{fig:operation}(b)).
Subsequently, horizontal transfer is performed at $K$~Hz to move each charge packet to its designated channel (right panel of Fig.~\ref{fig:operation}(b)).
By repeating these operations, each charge packet is transferred to its desired location.

Figure~\ref{fig:operation}(c) illustrates the waveform output phase.
The left panel of Fig.~\ref{fig:operation}(c) shows the state after the waveform loading phase is completed.
By waiting for a certain period, heat generated during the waveform loading phase can be allowed to thermalize before waveform output begins.
Waveform output is then performed as shown in the right panel of Fig.~\ref{fig:operation}(c).
During this phase, only horizontal transfer moves each charge packet toward the readout cell.
When a charge packet reaches the readout cell, a charge-to-voltage converter generates a voltage sample corresponding to amount of the stored charge.
By repeating this process, the stored sequence of charge packets is output as an analog voltage waveform.
To increase the effective output sampling rate, the proposed architecture employs time-interleaved output across multiple CCD channels, as shown in the right panel of Fig.~\ref{fig:operation}(c).
The target waveform is divided into multiple subsequences, each of stored in a separate CCD channel.
The output timings of the CCD channels are staggered relative to one another, and a selector switches sequentially among the channel outputs in the order of the target waveform samples.
When charge packets are transferred at a per-channel rate of $M~\mathrm{Hz}$ and $N$ CCD channels are interleaved, the combined output yields a baseband signal with an effective sampling rate of $MN$~samples per second (sps).

\section{Discussion}

\subsection*{Power Dissipation Estimate}

We estimate the heat load imposed on the cryogenic stage by the proposed CCD-based waveform-generation architecture.

\subsubsection{Reference Estimate Based on a Commercial CCD}

As a reference, we use a commercially available CCD, the Toshiba TCD2918BFG~\cite{toshiba_tcd2918bfg_jp}, to derive a conservative upper-bound estimate of the power required to drive the CCD transfer clocks. 
We then discuss how a purpose-built CCD for qubit control could reduce this dissipation.

The TCD2918BFG is a 10,680-pixel linear image sensor with a chip length of approximately 41~mm~\cite{toshiba_tcd2918bfg_jp}.
Its transfer-clock input capacitance is $C=100~\mathrm{pF}$, and the recommended clock voltage is $V=3.3~\mathrm{V}$.
This device is used here only as a reference point; a CCD designed specifically for qubit control would differ substantially in geometry, capacitance, operating voltage, and required storage length.

For an output sampling rate of $f_\mathrm{out}=500~\mathrm{MSps}$, which is typical for baseband signals used to control superconducting qubits~\cite{kurimoto2026microwave}, and with $N=10$ time-interleaved CCD channels, the per-channel clock frequency is $f_\mathrm{ch}=f_\mathrm{out}/N=50~\mathrm{MHz}$.
This frequency exceeds the TCD2918BFG's rated maximum clock frequency, 35~MHz and, therefore, the following calculation should be treated as an order-of-magnitude reference rather than as a specification of the commercial device.

Here, the charge-transfer pulse generator refers to the circuit that applies periodic voltage pulses to the CCD transfer electrodes to move charge packets along the CCD channel.
With $\Phi_\mathrm{elec}=4$ transfer electrodes per channel, the total power required to drive the transfer clocks can be estimated as
\begin{equation}
 P_\mathrm{clk} = C V^2 f_\mathrm{ch} \Phi_\mathrm{elec} N \approx 2.2~\mathrm{W}.   
\end{equation}

A major structural advantage of the proposed architecture is that the charge-transfer pulse generator can be located at room temperature, outside the cryogenic stage.
Only a fraction of the electrical power delivered through the wiring is dissipated inside the refrigerator.
We denote this fraction by $\alpha = R_\mathrm{cryo} / (R_\mathrm{line} + R_\mathrm{cryo})$, where $R_\mathrm{cryo}$ is the wiring resistance in the cryogenic stage, and $R_\mathrm{line}$ is the total line resistance between the room-temperature driver and the cryogenic load.
The corresponding cryogenic heat load associated with charge transfer is then $\alpha P_\mathrm{clk}$.
For example, assuming $\alpha=0.01$--$0.10$, the reference device would produce a cryogenic heat load of approximately $22$--$220~\mathrm{mW}$.

\subsubsection{Potential Reduction Through Design Optimization}

The large transfer-clock capacitance of the TCD2918BFG primarily reflects its imaging-oriented geometry.
The device contains 10,680 pixels per line and has a chip length of approximately 41~mm, which is far greater than what is required for qubit-control waveform generation.
For example, for a per-channel output rate of $50~\mathrm{MHz}$ and a waveform duration of $1~\mu\mathrm{s}$, only 50 stored samples are required per CCD channel.
If the transfer-clock capacitance scales approximately with the number of storage sites, this reduction in storage length yields
$100~\mathrm{pF}\times 50/10680 \simeq 0.47~\mathrm{pF}$.
Accounting for additional parasitic capacitance from wiring, contacts, and peripheral structures, we use $C=1~\mathrm{pF}$ as a representative target value for a purpose-built CCD waveform-generation device.

Using this representative optimized capacitance of $C=1~\mathrm{pF}$ while keeping the clock voltage at $V=3.3~\mathrm{V}$, the optimized transfer-clock power is estimated to be
$P_\mathrm{clk,opt}\approx 22~\mathrm{mW}$.
This value is approximately 100 times lower than the reference estimate based on the commercial CCD.
For $\alpha=0.01$--$0.10$, the corresponding cryogenic heat load is approximately $0.22$--$2.2~\mathrm{mW}$.
Note that whether this target capacitance can be achieved at $4~\mathrm{K}$ remains an open question and requires device-level characterization of a CCD structure specifically designed for cryogenic qubit-control waveform generation.

\section{Conclusion}

We have proposed an arbitrary waveform-generation architecture using CCDs for qubit control, in which waveform samples are stored as charge packets and read out sequentially as voltage samples. 
Time-interleaved operation across multiple CCD channels increases the effective output sampling rate, and the generated baseband waveforms are upconverted to microwave control signals for qubit operations. 
By storing waveform information as charge and reading it out on demand, the proposed architecture provides a distinct design direction for scalable cryogenic control systems, with the potential to reduce cryogenic interconnects and mitigate heat load compared with conventional approaches. 
Future work includes cryogenic device characterization and system-level power comparisons with existing qubit-control architectures.

%%%%%%%%%%%%%%%%%%%%%%%%%%%%%%%%%%%%%%%%%%%%%%%%%%%%%%%%%%%%%%%%%%%%%%%%%%%%%

\section*{Acknowledgment}

This work was based on results obtained from a project (Grant Number JPNP26008) subsidized by the New Energy and Industrial Technology Development Organization (NEDO), and supported by JST Moonshot R\&D (Grant Number JPMJMS256G).

%%%%%%%%%%%%%%%%%%%%%%%%%%%%%%%%%%%%%%%%%%%%%%%%%%%%%%%%%%%%%%%%%%%%%%%%%%%%%

% \bibliographystyle{IEEEtran}

\printbibliography

%\bibliography{IEEEexample}

\end{document}